\documentclass{elsart}

\usepackage{amssymb}
\usepackage{amsmath}
\begin{document}

\begin{frontmatter}

\title{Lower Bounds for Optimal Alignments of Binary Sequences}

\author{Cynthia Vinzant}

\address{Department of Mathematics, University of California, Berkeley 94720}

\ead {cvinzant@math.berkeley.edu}

\begin{abstract}
In parametric sequence alignment, optimal alignments of two sequences are computed as a function of the penalties for mismatches and spaces, producing many different optimal alignments. 
Here we give a $3/(2^{7/3}\pi^{2/3})n^{2/3} +O(n^{1/3} \log n)$ lower bound on the maximum number of distinct optimal alignment summaries of length $n$ binary sequences. This shows that the upper bound given by Gusfield et. al. is tight over all alphabets, thereby disproving the ``$\sqrt{n}$ conjecture".  Thus the maximum number of distinct optimal alignment summaries (i.e. vertices of the alignment polytope) over all pairs of length $n$ sequences is $\Theta(n^{2/3})$.
\end{abstract}

\end{frontmatter}

\section{Introduction and Notation}

Finding optimal alignments of DNA or amino acid sequences is often used in biology to measure sequence similarity (homology) and determine evolutionary history.  For a review of many problems relating to sequence alignment, see \cite{G book} and \cite{ASCB}. Here we deal with the question of how many different alignment summaries can be considered optimal for a given pair of sequences (though many different alignments may correspond to the same alignment summary). 

Given sequences $S$, $T$, an \textit{alignment} $\Gamma$ is a pair $(S', T')$ formed by inserting spaces, ``$-$", into $S$ and $T$. In each position, there is a \textit{match}, in which $S'$  and $T'$ have the same characters, a \textit{mismatch}, in which they have different characters, or a space in one of the sequences. 
Then for any alignment, we have an \textit{alignment summary} $(w, x,y)$, where $w$ is the number of 
matches, $x$ is the number of mismatches, and $y$ is the number of spaces in one of the sequences. Notice that $n = w+x+y$, where $n$ is the length of both sequences. Given a pair of sequences, the convex hull of all such points $(w,x,y)$ is called their \emph{alignment polytope}. 

We can score alignments by weighting each component. Since we have $w+x+y=n$, we can normalize so that the weight of $w$ is 1, the weight of $x$ is $-\alpha$ and the weight of $y$ is $-\beta$. Then \[score_{(\alpha, \beta)}(w,x,y) = w - \alpha x - \beta y.\] A sequence is \textit{optimal} if it maximizes this score. For biological relevance, we will only consider non-negative $\alpha$ and $\beta$, which penalizes mismatches and spaces. It is also possible to weight other parameters, such as \textit{gaps} (consecutive spaces) or mismatches between certain subsets of characters.   Here we will consider only the two parameter model described above. 

\begin{exmp} \label{ex1} For the sequences  111000 and 010110, we have an alignment \begin{equation*} 
\begin{matrix} 
- & 1 & - & 1&1& 0 &0 &0\\ 
0 & 1 & 0 & 1&1&1&-&-
\end{matrix} \qquad \end{equation*} 
which has 3 matches, 1 mismatch, and 2 spaces. So for a given $\alpha$ and $\beta$ the score of the this alignment would be $3 - \alpha - 2\beta$.
\end{exmp}

Any value of $\alpha$ and $\beta$ will give an optimal alignment.  Given $\alpha$ and $\beta$, we can use the Needleman-Wunsch algorithm to effectively compute optimal alignments \cite{NW} (for a review, see \cite[Ch. 2, 7]{ASCB}).  Unfortunately, different choices for $\alpha, \beta$ give different optimal alignments, leaving the problem of which weights to use. To resolve this,  Waterman, Eggert, and Lander proposed \emph{parametric alignment}, in which the weights $\alpha$, $\beta$ are viewed as parameters rather than constants \cite{beginnings}. Since alignments are discrete, this creates a partition of the $(\alpha, \beta)$ plane into \textit{optimality regions},  so that for each region $R$, there is an alignment that is optimal for all the points on its interior and $R$ is maximal with this property \cite{G}.  
Each optimality region is a convex cone in the plane \cite{G}, \cite[Ch. 8]{ASCB}.  Notice that because our scoring function is linear, the vertices of the alignment polytope are our optimal alignment summaries. Also, if we let $P_{xy}$ be the convex hull of all $(x,y)$ occurring in alignment summaries, then \[score_{(\alpha, \beta)} = w - \alpha x - \beta y = n - (\alpha +1)x - (\beta +1)y,\] since $n= w+x+y$. Thus the vertices of $P_{xy}$ will be those that minimize $(x,y)\cdot (\alpha+1, \beta+1)$ for some $(\alpha, \beta$), thus maximizing $score_{(\alpha, \beta)}$ and corresponding to optimal alignments \cite{ASCB}. From this we can see that the the decomposition of the $(\alpha, \beta)$ plane into optimality regions can be obtained by shifting the normal fan of $P_{xy}$ by $(-1, -1)$ \cite[Ch. 8]{ASCB}. The goal of parametric alignment is to find all these optimality regions with their corresponding optimal alignments. The Needleman-Wunsch algorithm is also an effective method of computing the alignment polytope of sequences (and thus  optimal alignments and the decomposition of the $(\alpha, \beta)$ plane) \cite{ASCB}. 

Gusfield et. al. showed that for two sequences of length $n$, the number of optimality regions of the $(\alpha, \beta)$  plane (equivalently the number of vertices in their alignment polytope) is $O(n^{2/3})$\cite{G}.  Indeed for larger dimensional models (say with $d$ free parameters), this bound was extended  to $O(n^{d-(1/3)})$  by Fern\'andez-Baca et. al. \cite{Baca2} and improved to $O(n^{d(d-1)/(d+1)})$ by Pachter and Sturmfels \cite{alg bounds}. For $d=2$, Fern\'andez-Baca et. al. refined this bound to $3(n/2\pi)^{2/3} + O(n^{1/3}\log (n))$ and showed it to be tight over an infinite alphabet \cite{Baca}. They also provide a lower bound of $\Omega(\sqrt{n})$ over a binary alphabet. Using randomly-generated sequences, Fern\'andez-Baca et. al. observed that the average number of optimality regions closely approximates $\sqrt{n}$. This led them to conjecture that, over a finite alphabet, the expected number of optimality regions is $\Theta(\sqrt{n})$\cite{Baca}.  The question remained of whether or not the upper bound of Gusfield et. al. was tight over a finite alphabet. For a discussion, see \cite[Ch. 8]{ASCB}, which conjectures that the maximum number of optimality regions induced by any pair of length-$n$ binary strings is $\Theta(\sqrt{n})$ \cite{ASCB}. Here we construct a counterexample to this conjecture, which together with the above upper bounds shows it instead to be $\Theta(n^{2/3})$. Our main theorem is that Gusfield's bound is tight for binary strings.

\begin{thm}[Main Theorem] The maximum number of optimality regions induced by binary strings of length $n$ is  $\Theta(n^{2/3})$. \end{thm}

Ideally, sequences would have few optimal alignments, making the ``best" one more apparent. While this result may not tell us about the expected number of optimal alignments (or be biologically relevant), it does provide a worst case scenario for sequence alignment and show that the bound from \cite{G} cannot be improved.  Luckily, the bound is still sublinear. Indeed parametric sequence alignment can be practical and has been achieved for whole genomes \cite{fly}. This paper is mainly motivated by \cite{Baca}, \cite{G}, and \cite{ASCB}.  We largely follow their notation and presentation. 

\section{Decomposing the $(\alpha, \beta)$ plane}

\subsection{Alignment Graphs}
\begin{figure} \begin{center}
\begin{picture}(45, 28)

\multiput(4,4)(2,0){7}{\line(0,1){12}}
\multiput(4,4)(0,2){7}{\line(1,0){12}}

\multiput(2.5,4.5)(0,2){3}{1}
\multiput(2.5, 10.5)(0,4){2}{0}
\put(2.5, 12.5){1}

\multiput (4.5, 16.5)(2,0){3}{1}
\multiput (10.5, 16.5)(2,0){3}{0}

\thicklines
\put(4,14){\line(1, -1){2}}
\put(6,10){\line(1,-1){6}}

\linethickness{0.5mm} 
\put(4,16){\line(0, -1){2}}
\put(6,12){\line(0,-1){2}}
\put(12,4){\line(1,0){4}}

\thinlines

\multiput(4,10)(.4,0){15}{\line(0,-1){6}}
\multiput(4, 14)(.4,0){15}{\line(0,-1){2}}

\multiput(10, 16)(.4,0){15}{\line(0,-1){2}}
\multiput(10, 12)(.4,0){15}{\line(0,-1){2}}

\put(31,0){\vector(0,1){22}} 
\put(25,6){\vector(1,0){18}}
\multiput(30.5,2)(0,4){5}{\line(1,0){1}}
\multiput(27,5.5)(4,0){4}{\line(0,1){1}}

\put(27,2){\circle*{0.7}}

\put(25,0){\line(1,1) {18}}
\put(26, 0){\line(1,2){11}}

\put(32,19){$\Gamma_1$}
\put(35,14){$\Gamma_2$}
\put(38, 9){$\Gamma_3$}

\put(29,20){$\beta$}
\put(42,4){$\alpha$}

\end{picture} \end{center}
\label{fig: example}

\caption{\textbf{(Left)}:Above is the path corresponding to the alignment of 111000 and 010111 given in example \ref{ex1}, (-1-11000, 010111- -). The shaded regions denote possible matches. \;\;\;\;\;\;\;\;\;\;\;\;\;\;\;\;\;\;\;\;\;\;\;\;\;\;\;\;\;\; \;\;\;\;\;\;\;\;\;\;\;\;\;\;\;\;\;\;\;\;\;\;\;\;\;\;\;\;\;\;  \;\;\;\;\;\;\;\;\;\;\;\;\;\;\;\;\;\;\;\;\;\;\;\;\;\;\;\;\;\;\;\;\;\;\;\;\;\;\;\;\;\;\;    \textbf{(Right)}: Here are the optimality regions of the $(\alpha, \beta)$ plane induced by the sequences 111000 and 010111. The alignments optimized in each region are $\Gamma_1 = (111000, 010111)$, $\Gamma_2 = $(-1-11000, 010111- -), and \;\;\;\;\;\;\;\;\;\;\;\;\;\;\;\;\;\;\; \;\;\;\;\;\;\;\;\;\;\;\;\;\;\;\;\;\;\;\;\;\;\;\;\;\;\;\;\;\; $\Gamma_3 = $(-1-11-000, 010111- - -).} \end{figure}
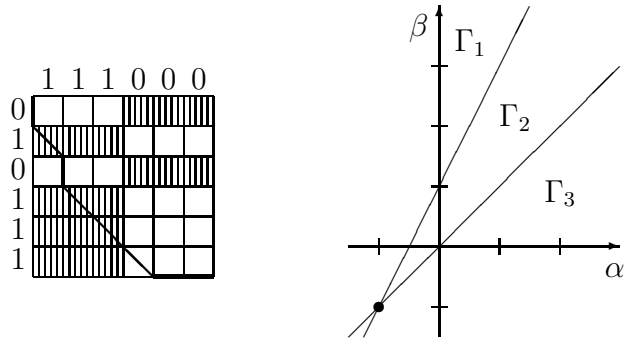

  We can represent every alignment of two length-$n$ sequences as a path through their \textit{alignment graph}. The graph can be thought of as an $(n+1) \times (n+1)$ grid, with rows and columns numbered consecutively from top to bottom (left to right), from 0 to $n$ \cite{Baca}. An \textit{alignment path} is a path on these vertices, starting at $(0,0)$, ending at $(n,n)$, and only moving down, right or diagonally down and to the right.  Each path corresponds to a unique alignment. In this path, a move down (or left) corresponds to a space in the first (or second) sequence, and a diagonal move corresponds to a match or mismatch (depending on the characters). See Figure~\ref{fig: example} for the alignment graph of our above example alignment.

\subsection{Optimality regions}
Gusfield et. al. observed that the boundaries between optimality regions in the $(\alpha, \beta)$ plane must be lines passing through the point $(-1,-1)$.

\begin{lem} [Gusfield et. al., \cite{G}] All optimality regions on the $(\alpha, \beta)$ plane are semi-infinite cones, and are delimited by lines of the form $\beta = c + (c+1)\alpha$ for some constant $c$. 
\end{lem}

In general, a boundary between two optimality regions consists of the $(\alpha, \beta)$ for which the optimal sequences from each region have equal, optimal scores. Since 
\[ score_{(-1, -1)}(w,x,y) = w +x +y \equiv n, \]
 for every $w, x, y$, each such line (specifically these boundary lines) must pass through the point $(-1,-1)$.  
 They also note that all of these boundary lines must intersect the non-negative $\beta$-axis because none of them cross the positive $\alpha$-axis \cite{G}. This comes from observing that in any alignment, we can change a mismatch to a space (in each sequence) without affecting the number of matches. Thus all along the line $\beta =0$, the optimal alignment will have the maximum number of matches possible, without regard to spaces (since those are not penalized).  So no boundary line can separate the nonnegative $\alpha$-axis into distinct optimality regions.  Since all boundary lines must pass through the point $(-1,-1)$ and cannot intersect the positive $\alpha$-axis, we indeed have that
 
\begin{lem} [Gusfield et. al., \cite{G}]  Each of the optimality regions must have nontrivial intersection with the non-negative $\beta$-axis.  That is, for any path $\Gamma$ that is optimized by some $(\alpha, \beta)$, there must be some $\beta '$ so that $\Gamma $ is optimized by $(0, \beta')$.
 \end{lem}
 
 This allows us to restrict our attention to optimality regions on the $\beta$-axis. Then boundary regions are just points, $(0,\beta)$, for which consecutive optimal alignments have optimal $score_{(0,\beta)}$. Note that alignments with summaries $(w_1, x_1, y_1)$ and $(w_2, x_2, y_2)$ will have equal $score_{(0,\beta)}$ when \[w_1 - \beta y_1 = w_2 - \beta y_2,\] 
 meaning that \[\beta = \frac{\Delta w}{\Delta y} := \frac{w_2-w_1}{y_2-y_1}.\]
 
In order to find different optimality regions, we will find distinct $\frac{\Delta w}{\Delta y}$ forming boundary points on the $\beta$-axis.

\section{The Lower Bound}
For each $2 \leq r$, define $F_r$ as \begin{center} $F_r:= \{\frac{a}{b}\leq 1 \; : \; \frac{a}{b}$ is reduced and $a+b = r\}$. \end{center} 
Since $a/b$ is reduced and $a+b=r$, $a$ and $b$ must be relatively prime to $r$. Then each number relatively prime to $r$ will show up exactly once (in either the numerator or the denominator), so $|F_r| = \phi(r)/2$ for $r >2$ where $\phi$ is the Euler totient function, and $|F_2| = |\{1/1\}| =1$.

Let \[\mathcal{F}_q = \bigcup_{r=2}^q F_r,\]
giving us $|\mathcal{F}_q| = \frac{1}{2}\sum_{r=3}^{q}\phi(r) +1$. \\ \vspace{2mm}\\
Fixing $q$, let $a_1/b_1 < a_2/b_2 < \ldots  < a_m/b_m=1$ be the elements of $\mathcal{F}_q$. We're going to construct two sequences of length $n = 4 \sum_kb_k $, $S = s_1s_2 \hdots s_n$ and $T=t_1t_2\hdots t_n$. Since $b_k < a_k+ b_k$, this gives us 
\[ n = 4\sum_{k=1}^m b_k < 4 \sum_{k=1}^m (a_k + b_k) = 4\sum_{r=2}^s r |F_r| = 2 \sum_{r=2}^s r \phi(r).\]

\subsection{The Sequences}
Let's construct the first sequence, $S$. To start, let the first $b_1+a_1$ elements of $S$ be 0, followed by $b_1-a_1$ 1's. 
Then repeat for $k>1$ (i.e. next place $b_2+a_2$ 0's followed by $b_2-a_2$ 1's). Notice that for each $a_k/b_k \in \mathcal{F}_q$, we use $(b_k+a_k)+ (b_k-a_k)=2b_k$ places.  To get the second half of the sequence, take the reverse complement of the first half (reflecting it and switching all the 1's and 0's). So 
 \[ S = 0^{b_1+a_1}1^{b_1 - a_1} 0^{b_2+a_2}\hdots  0^{b_m+a_m}1^{b_m - a_m} \;\; 0^{b_m-a_m}1^{b_m + a_m}\hdots 0^{b_1-a_1}1^{b_1 + a_1}.\]

More formally, define \[i(r) = \sum_{k=1}^r 2b_k \;\;\;\;\; \text{         and         } \;\;\;\;\;  j(r) = \sum_{k=r}^m 2b_k.\]
(So $n = 2i(m) = 2j(1)$). Then \begin{equation*} 
s_{i(r-1) +k} = \left\{ 
\begin{array}{rl} 
0 & \text{for } 1 \leq k \leq b_r +a_r\\ 
1 & \text{for } b_r+a_r +1\leq k \leq 2b_r
\end{array} \right. 
\end{equation*} 
and 
\begin{equation*} 
s_{\frac{n}{2}+j(r+1)+k} = \left\{ 
\begin{array}{rl} 
0 & \text{for } 1 \leq k\leq b_r -a_r\\ 
1 & \text{for } b_r-a_r +1\leq k \leq 2b_r.
\end{array} \right. 
\end{equation*} 

The second sequence, $T$, will just be $n/2$ 1's followed by $n/2$ 0's, that is,

\begin{equation*} 
t_k = \left\{ 
\begin{array}{rl} 
1 & \text{for } 1 \leq k\leq n/2\\ 
0 & \text{for } n/2 +1 \leq k \leq n.
\end{array} \right. 
\end{equation*}

\begin{exmp} \label{ex: main}For $q = 4$, $\mathcal{F}_4 = \{1/3, 1/2, 1/1 \}$. Then $n = 4(3+2+1)= 24$. Our sequences are \[S = 000011000100\; 110111001111\]  \[T = 111111111111\; 000000000000\] 
\end{exmp}

\subsection{The Alignment Paths}
We are going to construct $m+1$ alignment paths, $\Gamma_{m+1}, \Gamma_m, \hdots, \Gamma_1$. 
Let $\Gamma_{m+1} $ be the path along the main diagonal (corresponding to the alignment with no spaces).  To get $\Gamma_r$, align the first $j(r) =  \sum_{k=r}^m 2b_k$ 0's of $S$ with spaces and align its remaining elements without spaces, ending by aligning the last $j(r)$ 0's of $T$ with spaces. 
 
\begin{figure} \begin{center}
\begin{picture}(56,51)
\multiput(5,0)(2,0){25}{\line(0,1){48}}
\multiput(5,0)(0,2){25}{\line(1,0){48}}
\multiput(3.5,.5)(0,2){4}{1}
\multiput(3.5,8.5)(0,2){2}{0}
\multiput (3.5,12.5)(0,2){3}{1}
\put (3.5, 18.5){0}
\multiput(3.5,20.5)(0,2){2}{1}
\multiput (3.5, 24.5)(0,2){2}{0}
\put (3.5,28.5){1}
\multiput (3.5,30.5)(0,2){3}{0}
\multiput (3.5,36.5)(0,2){2}{1}
\multiput (3.5,40.5)(0,2){4}{0}

\multiput (5.7,48.5)(2,0){12}{1}
\multiput (29.7,48.5)(2,0){12}{0}

\thicklines
\put(5,48){\line(1, -1){48}}
\put(5,44){\line(1,-1){44}}
\put(5,40){\line(1,-1){4}}
\put(9,32){\line(1,-1){32}}
\put(9,30){\line(1,-1){2}}
\put(11, 24){\line(1,-1){4}}
\put(15, 18){\line(1,-1){6}}
\put(21, 8){\line(1,-1){8}}

\linethickness{0.5mm} 
\put(5,48){\line(0, -1){8}}
\put(9,36){\line(0,-1){6}}
\put(11,28){\line(0,-1){4}}
\put(15, 20){\line(0,-1){2}}
\put(21, 12){\line(0,-1){4}}
\put(29,0){\line(1,0){24}}

\thinlines

\multiput(29,48)(.4,0){60}{\line(0,-1){8}}
\multiput(29,36)(.4,0){60}{\line(0,-1){6}}
\multiput(29,28)(.4,0){60}{\line(0,-1){4}}
\multiput(29,20)(.4,0){60}{\line(0,-1){2}}
\multiput(29,12)(.4,0){60}{\line(0,-1){4}}

\multiput(5,40)(.4,0){60}{\line(0,-1){4}}
\multiput(5,30)(.4,0){60}{\line(0,-1){2}}

\multiput(5,24)(.4,0){60}{\line(0,-1){4}}
\multiput(5,18)(.4,0){60}{\line(0,-1){6}}
\multiput(5,8)(.4,0){60}{\line(0,-1){8}}

\put(29,0){\circle*{0.6}} 
\put(29,12){\circle*{0.6}} 
\put(29,20){\circle*{0.6}} 
\put(29,24){\circle*{0.6}}
\end{picture} \end{center}
\label{fig: gammas}
\caption{The alignment graph for $\Gamma_4, \Gamma_3, \Gamma_2, \Gamma_1$ (top to bottom) from the example above. The shaded regions denote possible matches. Note that for $q=4$, $m=|\mathcal{F}_4| =3$.}
\end{figure}
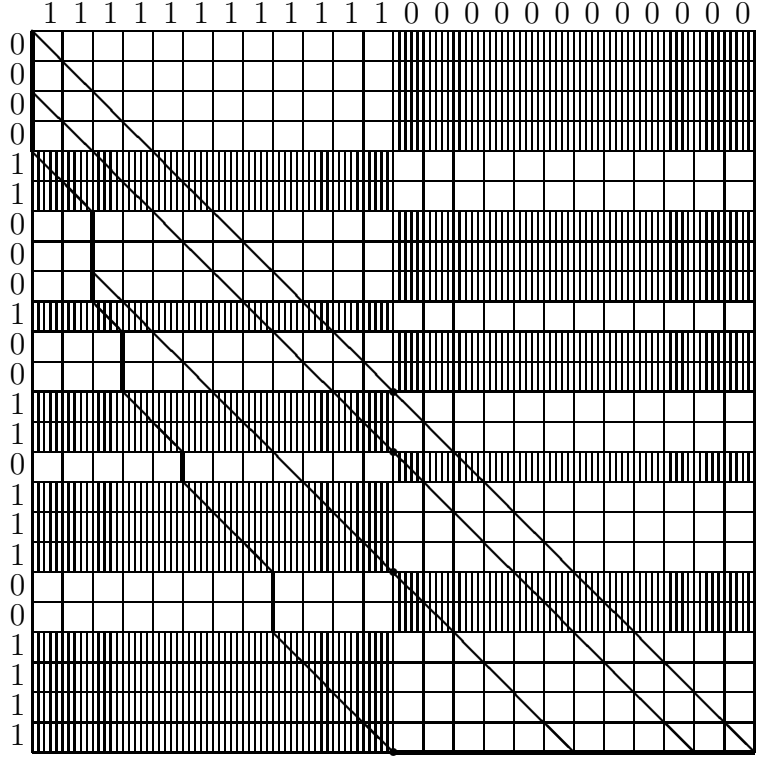

Note that because there are $n/2$ 1's in both $S$ and $T$, we'll have enough room to do this. In fact, in the last alignment, $\Gamma_1$, all the 1's of $S$ will be matched with all the 1's of $T$. See Figure~\ref{fig: gammas} for the graphs of the optimal alignments of our example.

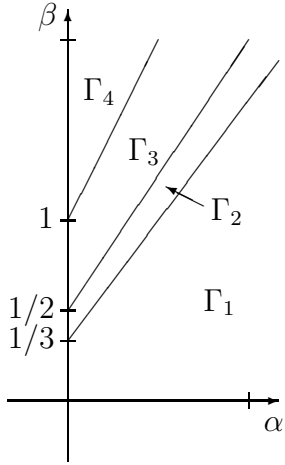
\begin{figure} \begin{center}
\begin{picture}(20, 32)

\put(4,0){\vector(0,1){30}} 
\put(0,4){\vector(1,0){18}}

\put(3.5,8){\line(1,0){1}}
\put(3.5,10){\line(1,0){1}}
\put(3.5, 16){\line(1,0){1}}
\put(3.5, 28){\line(1,0){1}}
\put(16, 3.5){\line(0,1){1}}

\put(0,7.5){1/3}
\put(0,9.5){1/2}
\put(2,15.5){1}

\put(2,29){$\beta$}
\put(17, 2){$\alpha$}

\put(4,8){\line(3,4){14}}
\put(4,10){\line(2,3){12}}
\put(4,16){\line(1,2){6}}

\put(5,24){$\Gamma_4$}
\put(8,20){$\Gamma_3$}
\put(13.5,16){$\Gamma_2$}
\put( 13,10){$\Gamma_1$}

\put(13,17){\vector(-2,1){2.5}}

\end{picture} \end{center}
\label{fig: ab decomp}
\caption{Here is the decomposition of the $(\alpha, \beta)$ plane given by the sequences in Example \ref{ex: main}.  Each optimality region is labeled with the alignment path $\Gamma_r$ that it optimizes. }
\end{figure}

 \subsection{Alignment Scores} Let $w_r^1$ denote the number of matching 1's in $\Gamma_r$ and similarly $w_r^0$ denote the number of matching 0's in $\Gamma_r$, with $w_r$ being the total number of matches. Note that \[w_r^1 - w_{r+1}^1 = b_r +a_r \;\; \text{ and }\;\;w_r^0 - w_{r+1}^0 = - (b_r-a_r).\] Since $w_r = w_r^1+w_r^0$, we have that \[w_r - w_{r+1} = (b_r+a_r) - (b_r-a_r) = 2a_r.\]
 Let $y_r$ denote the number of spaces in $\Gamma_r$ (which equals $j(r)$).  Then \[y_r - y_{r+1} = j(r) - j(r+1) = 2b_r.\]
 
 Putting these together, we get that for every $r$, \begin{equation} \frac{\Delta w_r}{\Delta y_r} := \frac{w_r - w_{r+1}}{y_r - y_{r+1}} = \frac{a_r}{b_r}.\label{a/b achieved}\end{equation}
 
 \subsection{Optimality}
 We need to show that each of these paths is optimal for distinct optimality regions, which will be accomplished by the next two lemmata. 
 
\begin{lem}\label{optimal}  Let $\Gamma$ be any alignment of $S$ and $T$. Then for any $\beta \geq 0$, there is  some $\Gamma_r$ so that  $score_{(0,\beta)}(\Gamma_r) \geq score_{(0,\beta)}(\Gamma)$.
\end{lem}
\begin{pf}
Say that $\Gamma$ has alignment path $\sigma$ and alignment summary $(w,x,y)$. 
Let the coordinates of the alignment graph be $(t,s)$, with $(0,0)$ starting in the upper left corner. Say that $(n/2, n/2+k)$ is the first time $\sigma$ meets the vertical line $t=n/2$.

Because of the symmetry of our sequences, we can take $k$ to be nonnegative (meaning that $\sigma$ hits the line $t=n/2$ below or at $s=n/2$). If $\sigma$ has $k<0$, we can rotate our picture $180^o$  to get another alignment path with the same summary and $k \geq 0$. 

So suppose $k \geq 0$ and take $r$ so that $j(r+1) < k \leq j(r)$.\\
 \vspace{2mm} \\  
\emph{(Case 1:} $k-j(r+1) \leq b_r - a_r$\emph{)}.

Since there are only $w_{r+1}^1$ 1's above $s=n/2+k$, we have $w^1\leq w^1_{r+1}$. Similarly, there are at most $w_{r+1}^0$ 0's below $s=n/2+k$, so $w^0 \leq w_{r+1}^0$. 
Furthermore, by going through the point $(n/2, n/2+k)$, $\sigma$ must have at least $k$ spaces, so $y \geq k \geq j(r+1) = y_{r+1}$. 
Putting these together gives that for any $\beta \geq 0$, 
\[ score_{(0,\beta)}(\Gamma_{r+1}) - score_{(0,\beta)}(\Gamma) = (w_{r+1} - w) - \beta (y_{r+1} - y) \geq  0 .\]
Intuitively, $\Gamma$ can have at most as many matches and must have at least as many spaces as $\Gamma_{r+1}$, and thus cannot have a higher score. \\ 
\vspace{2mm}\\
\emph{(Case 2: $k-j(r+1) > b_r - a_r$ and $\beta \leq 1$)}

There are $w_r^0$ 0's in $S$ below $s= n/2+k$, so we have $w^0 \leq w^0_r$.  In addition to the $w_{r+1}^1$ 1's in $S$ above $s=n/2 + j(r+1)$, there are another $k-j(r)+(b_r+a_r)$ 1's in $S$ between $s=n/2+j(r+1)$ and $s=n/2+k$. So 
\[w^1 \leq w_{r+1}^1 +k-j(r)+(b_r +a_r) = w_r^1 + k -j(r),\]
since $w_{r+1}^1 + (b_r +a_r)= w_r^1$. Thus \begin{equation}w = w^0 +w^1 \leq w^0_r + w^1_r +k - j(r) = w_r +k -j(r).\label{w_r}\end{equation}  As is case 1, we have that $y\geq k$, so \begin{align} 
score_{(0,\beta)}(\Gamma_r) - score_{(0,\beta)}(\Gamma) &= (w_r - w) - \beta(y_r - y)  \nonumber\\ 
&\geq (j(r) - k) - \beta(j(r) - k)  \tag{by \eqref{w_r}}  \nonumber  \\ 
& \geq 0  \tag{as $\beta \leq 1$} \nonumber 
\end{align}\\
 \emph{(Case 3: $k-j(r+1) > b_r - a_r$ and $\beta > 1$)}
 
 We'll show that $score_{(0,\beta)}(\Gamma_{m+1}) \geq score_{(0,\beta)}(\Gamma)$. Remember that $\Gamma_{m+1}$ is the alignment with no spaces ($y_{m+1} = 0$), corresponding to the main diagonal of the alignment graph. Note for any $r$, \begin{equation} \label{w_m+1} w_r  =  w_{m+1}+\sum_{k=r}^m 2a_k,\end{equation}
so using equation \eqref{w_r} from case 2, we get
\[w_{m+1} - w \geq j(r)-k - \sum_{k=r}^m 2a_k. \]
As in previous cases, $y \geq k$. Then, \begin{align} 
score_{(0,\beta)}(\Gamma_{m+1}) - score_{(0,\beta)}(\Gamma)
& = (w_{m+1} - w) - \beta(y_{m+1} -y) \nonumber\\
& \geq j(r)-k - \sum_{k=r}^m 2a_k  + \beta k \tag{by \eqref{w_m+1}} \nonumber\\ 
&\geq   j(r) - \sum_{k=r}^m 2a_k \tag{as $\beta >1$} \nonumber  \\ 
& = \sum_{k=r}^m 2b_k  - \sum_{k=r}^m 2a_k \nonumber \\
& \geq 0. \nonumber
\end{align}
\end{pf} Lemma \ref{optimal} tells us that any optimality region has one of the $\Gamma_r$ as an optimal alignment. Now we need to check that each $score_{(0,\beta)}(\Gamma_r)$ is optimized by a different region. 
To see this, we use equation \eqref{a/b achieved} and following lemma.
\begin{lem} [Fern\'andez-Baca, et. al., \cite{Baca}]  Let $\Gamma_1, \Gamma_2, \hdots, \Gamma_q$ be paths in the alignment graph. Assume $score(\Gamma_i)=w_i - \beta y_i$, where $y_1 > y_2 > \hdots > y_q$. Let $\beta_0 = 0, \beta_q = \infty$, and for $r = 1, \hdots, q-1$, $\beta_r = (w_r - w_{r+1})/(y_r - y_{r+1})$. Suppose $\beta_0 < \beta_1 < \hdots < \beta_q$. Then for $\beta \in (\beta_{r-1}, \beta_r)$ and $p \neq r$, $score_{(0,\beta)}(\Gamma_r) > score_{(0,\beta)}(\Gamma_p)$. 
\end{lem}

So each of the $\Gamma_r$ do indeed represent each of the different optimality regions on the $\beta$-axis, and thus in the $(\alpha, \beta)$ plane. 

\subsection{The Actual Lower Bound}

\begin{thm}
The maximum number of optimality regions induced by any pair of length-$n$ sequences is $\Omega(n^{2/3})$.
\end{thm}
\begin{pf}
Above we have constructed sequences of length $n \leq 2 \sum_{r=2}^q r\phi(r)$ that gave $m = \frac{1}{2}\sum_{r=2}^q \phi(r)$ optimality regions. 
From analytic number theory, as calculated in \cite{Baca},
\[m= \frac{1}{2}\sum_{r=3}^q \phi(r) +1 = \frac{3}{2 \pi^2}q^2 +O(q \log q),\]
and
\[n \leq 2\sum_{r=2}^q r\phi(r) = \frac{4}{\pi^2}q^3 + O(q^2\log q). \]
Then $q \geq (\frac{\pi^2 n}{4})^{1/3} +O(\log n)$, meaning

\begin{align}
m= \frac{1}{2} \sum_{r=3}^q \phi(r) +1
& \geq \frac{3}{2\pi^2}\left((\frac{\pi^2 n}{4})^{1/3}\right)^2 +O(n^{1/3} \log n) \nonumber\\
& = \frac{3}{2^{7/3}\pi^{2/3}}n^{2/3} +O(n^{1/3} \log n). \nonumber
\end{align}
\end{pf}

With the upper bounds from \cite{G} and \cite{Baca}, this gives

\begin{cor} The maximum number of optimality regions over all pairs of length-$n$ sequences is $\Theta(n^{2/3})$, and more specifically is between  $\frac{3}{2^{7/3}\pi^{2/3}}n^{2/3} +O(n^{1/3} \log n)$ and  $\frac{3}{(2\pi)^{2/3}}n^{2/3} +O(n^{1/3} \log n)$. \end{cor}

It's unclear whether the current bounds on optimality regions for scoring with $d>2$ parameters, $O(n^{d(d-1)/(d+1)})$, are also tight or whether better upper bounds exist. Another interesting open question (perhaps with more practical relevance) is the order of the expected number of optimality regions, rather than the maximum.

\section{Acknowledgements}
Thanks to Lior Pachter for his advice and suggestion of this problem.  This paper came out of his class at U.C. Berkeley, ``Discrete Mathematics for the Life Sciences", in the spring of 2008. Thanks also to Bernd Sturmfels and Peter Huggins for their useful suggestions.

\end{document}